\documentclass{vgtc}                          

\ifpdf
  \pdfoutput=1\relax                   
  \usepackage{graphicx}                
  \DeclareGraphicsExtensions{.pdf,.png,.jpg,.jpeg} 
\else
  \usepackage{graphicx}                
  \DeclareGraphicsExtensions{.eps}     
\fi%

\graphicspath{{figures/}{pictures/}{images/}{./}} 

\usepackage{microtype}                 
\PassOptionsToPackage{warn}{textcomp}  
\usepackage{textcomp}                  
\usepackage{mathptmx}                  
\usepackage{times}                     
\usepackage{cite}                      
\usepackage{tabu}                      
\usepackage{booktabs}                  
\usepackage{tabularx}
\usepackage{balance}
\usepackage{float}

\onlineid{0}

\vgtccategory{Research}

\vgtcinsertpkg

\title{Effects of data distribution and granularity on color semantics for colormap data visualizations}

\author{Clementine Zimnicki\thanks{e-mail: zimnicki@wisc.edu}\\ %
        \scriptsize  University of Wisconsin--Madison%
\and Chin Tseng\thanks{e-mail: chint@cs.unc.edu}\\ %
     \scriptsize University of North Carolina, Chapel Hill%
\and Danielle Albers Szafir\thanks{e-mail: danielle.szafir@cs.unc.edu}\\ %
    \scriptsize UNC, Chapel Hill%
\and Karen B Schloss\thanks{e-mail: kschloss@wisc.edu}\\ %
    \scriptsize UW--Madison%
     }

\teaser{
  \centering
  \includegraphics[width=1.0\linewidth]{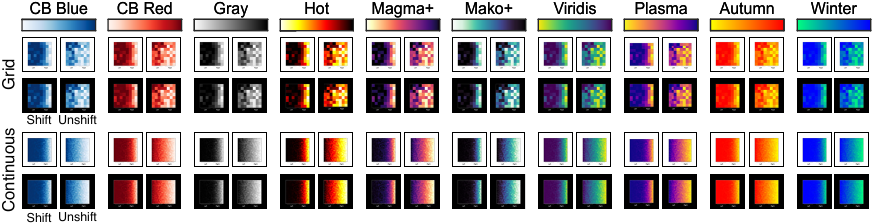}
  \caption{In this study, participants were asked which side of a colormap data visualization (left/right) represented data of greater magnitude. We applied 10 color scales to each of 8 conditions for each colormap: 2 granularity levels (grid/continuous) $\times$ 2 background colors (black/white) $\times$ 2 two data distributions (shifted or unshifted).}
  \label{fig:colormaps}
}

\abstract{To create effective data visualizations, it helps to represent data using visual features in intuitive ways. When visualization designs match observer expectations, visualizations are easier to interpret. Prior work suggests that several factors influence such expectations. For example, the dark-is-more bias leads observers to infer that darker colors map to larger quantities, and the opaque-is-more bias leads them to infer that regions appearing more opaque (given the background color) map to larger quantities. Previous work suggested that the background color only plays a role if  visualizations appear to vary in opacity. The present study challenges this claim.  
We hypothesized that the background color \textit{would} modulate inferred mappings for colormaps that should not appear to  vary in opacity (by previous measures) if the visualization appeared to have a ``hole'' that revealed the background behind the map (hole hypothesis). We found that spatial aspects of the map contributed to inferred mappings, though the effects were inconsistent with the hole hypothesis. Our work raises  new questions about how spatial distributions of data influence color semantics in colormap data visualizations. 
} 

\CCScatlist{
  \CCScatTwelve{Visual reasoning}{Information visualization}{Colormap data visualizations}{Color cognition}
}

\begin{document}


\firstsection{Introduction} \label{sec:intro}
\maketitle

When creating data visualizations, it is helpful to represent data in a way that is intuitive to observers. These intuitions stem from observer expectations about how visual features should map to concepts, called inferred mappings \cite{cuff1973, mcgranaghan1989, lin2013, setlur2016, schloss2018, schloss2019, schloss2021, mukherjee2022, schoenlein2023, soto2023}. Understanding inferred mappings is crucial for effective visual communication: when inferred mappings match the encoded mapping specified by the visualization designer, the visualization becomes easier to interpret \cite{tversky2002, norman2013, hegarty2011, tversky2011, lin2013, schloss2021, mukherjee2022, schoenlein2023}. 

In this paper, we focus on inferred mappings for colormap data visualizations (``colormaps'' for short), which represent
quantity using a gradation of color (``color scale'').\footnote{Various terms are used for visualizations that represent continuous data using gradations of color. Here, ``color scale'' refers to gradations of color used to construct a colormap (also known as ``ramps'' \cite{smart2019}, and ``colormap'' refers to a data visualization that represents quantities using gradations of color, such as in maps of weather patterns or visualizations of neural activity.}
To design colormaps that match observers' inferred mappings, a key consideration is deciding which endpoint of the color scale should map to larger quantities in the data. We focus solely on color scales that increase monotonically in lightness. We do not consider divergent color scales (light and dark at both endpoints)
or rainbow color scales \cite{ware1988, spence1999, borland2007, moreland2009, reda2021}. For color scales that vary monotonically in lightness, previous work suggests that several factors combine to 
influence whether observers infer that darker vs. lighter colors map to larger quantities.  

\textbf{Dark-is-more bias.} The dark-is-more bias leads to inferences that darker colors map to larger quantities \cite{cuff1973, mcgranaghan1989, schloss2019, sibrel2020}. Cuff \cite{cuff1973} provided early evidence for this: when asked to interpret colormaps with no legends, participants inferred that darker colors represented ``more.'' McGranaghan \cite{mcgranaghan1989} investigated whether the dark-is-more bias was actually a special case of a contrast-is-more bias. If so, the dark-is-more effect observed when the background is light should reverse when the background is dark. McGranaghan \cite{mcgranaghan1989} found that participants consistently inferred darker colors represented more on both light and dark backgrounds, though less so on the dark backgrounds. This challenged the notion that the dark-is-more bias was a special case of a contrast-is-more bias. But, contrast effects have been observed in other domains, such as in visual search \cite{rosenholtz2004}.

\textbf{Opaque-is-more bias.} Schloss et al. \cite{schloss2019} studied the effects of background for colormaps made using different color scales. They identified the opaque-is-more bias, which leads to the inference that regions appearing more opaque map to larger quantities (i.e., dark colors on light background and light colors on dark background). The strength of this bias (and effect of background lightness) depended on the degree to which colormaps appeared to vary in opacity (quantified using Opacity Variation Index). The strength of apparent opacity variation depends on how closely the color scale interpolates linearly with the background color (``value-by alpha'' maps \cite{roth2010}).

When both dark-is-more and opaque-is-more biases are activated, they work together on light backgrounds (darker regions appear more opaque) and conflict on dark backgrounds (lighter regions appear more opaque). Under such conflicts, the opaque-is-more bias can reduce, or even override the dark-is-more bias, resulting in inferences that lighter colors map to more \cite{schloss2019, bartel2021}. Critically, when colormaps did not appear to vary in opacity, background lightness had no effect on inferred mappings (i.e., no contrast-is-more bias).

\textbf{Hotspot-is-more bias.} Schott \cite{schott2010} suggested people expect regions in hotspots---concentric rings of data like those found in weather maps---map to more. Sibrel et al.\cite{sibrel2020} tested whether a hotspot-is-more bias exists, and if it could override the dark-is-more bias. They found that hotspot-is-more and dark-is-more biases worked together when hotspots were dark but conflicted when hotspots were light; under such conflicts, the dark-is-more bias dominated over the hotspot-is-more bias (leading to inferences that darker colors mean more) \cite{sibrel2020}. However, when hotspots were highly salient, the hotspot-is-more bias dominated---observers inferred that lighter colors mapped to more. Thus, spatial structure can impact inferred mappings.

\textbf{Direct associations.} The previous biases focused on perceptual factors, but another factor concerns the direct color-concept associations with the data domain (e.g., foliage, wildfire, sunshine) \cite{samsel2017, schoenlein2023}. When lighter colors in the colormap are strongly associated with the data domain (e.g., light yellow is strongly associated with sunshine), those direct associations can override the dark-is-more bias \cite{schoenlein2023}. 

\section{Motivation}
From previous work, it follows people will infer darker colors map to larger quantities, regardless of background lightness, if the following conditions are met: (1) the color scale does not appear to vary in opacity, (2) the colormap does not have salient light-colored hotspots, and (3) the colormap does not represent data about something with strong direct associations with light colors within the colormap.

In this study we considered that cases might exist in which these conditions are met, yet observers will infer that lighter colors map to more. This idea stems from Bartel et al. \cite{bartel2021}, who investigated inferences about the meaning of colors in Venn diagrams, which are systems for visually representing logical propositions \cite{venn1881}. Traditionally, shaded regions in Venn diagrams indicate ``non-existence'' of an entity represented by that region \cite{venn1881, shin1994}. However, Bartel et al. \cite{bartel2021} proposed the \textit{hole hypothesis}, which predicts people infer the opposite: regions appearing as ``holes'' in Venn diagrams map to non-existence. Regions appear as holes when their surface properties (e.g., color, texture) match the surface properties of the background of the display, resulting in the appearance that these regions are ``owned'' by the background \cite{nelson2001, peterson2003, bertamini2006}.  Supporting the hole hypothesis and contrary to Venn diagram conventions, Bartel et al. \cite{bartel2021} found that participants interpreted Venn diagrams consistently with regions appearing as holes that represented non-existence.

\begin{figure}[tb]
    \centering
    \includegraphics[scale=.95]{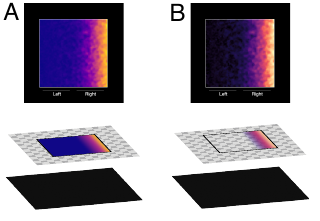} 
    \vspace{-5mm}
    \caption{Shifted colormaps created using Plasma/Magma+ \cite{hunter2007} color scales. (A) should not elicit the percept of a hole because the large dark area does not match the background. (B) should elicit the percept of a hole, as the large dark area matches the background \cite{nelson2001}.}
    \label{fig:hole}
    \vspace{-5mm}
\end{figure}

We hypothesized that if a colormap has a large black region on a black background, the region will appear like a hole, and this could activate the opaque-is-more bias even if the color scale used to construct the colormap does not appear to vary in opacity (hole hypothesis). For example Fig. \ref{fig:hole} shows two colormaps; Fig. \ref{fig:hole}A does not appear to have a hole as none of the colors match the background, whereas (Fig. \ref{fig:hole}B) appears to have a hole because the large black region matches the background. We focus here on \textit{perceptual holes} as described by \cite{nelson2001}, and distinguish these from regions indicating a lack of data (e.g., \cite{zhao2022}). If the hole hypothesis is supported, the probability of responding that darker colors represent larger values (``dark-more responses'') would be reduced for maps presented on a black background compared to on a white background. A sufficiently strong effect could override the dark-is-more bias, leading participants to make \textit{light-more} responses for colormaps on a black background. Thus, we would identify a condition in which people infer light is more, while the aforementioned conditions are met: (1) the color scale does not appear to vary in opacity, (2) the colormap does not have salient light-colored hotspots, and (3) the colormap does not represent data about something with strong direct associations with light colors within the colormap.

To test this hypothesis, we conducted an experiment assessing inferred mappings for colormaps that varied in \textit{data distribution}, background, and color scale, to create colormaps varying in how much they appeared to have holes (Fig. \ref{fig:colormaps}). We also varied \textit{granularity} (grid or continuous) to test if effects might be stronger for colormaps with smooth gradations rather than sharp edges.

\textbf{Contributions.} Our results make the following contributions.
First, we identified spatial factors, data distribution and granularity, contributing to inferences about the meaning of colors in colormap data visualizations. Second, we found that these factors were modulated by color scale and background, but not in ways we would expect based on prior work. Though our results raise unresolved questions concerning the role of spatial factors in inferred mappings, we have shown that it is crucial to account for the spatial structure of data when considering color semantics in data visualizations.

 \section{Methods}\label{sec:Methods}
\textbf{Participants.} Our target sample size was $n=1600$ ($n=20$ per condition). This sample size was determined by simulating expected error rates and standard deviation based on results from a study with a comparable design. We randomly sampled responses with replacement for $n=5-22$ in each condition and calculated the standard deviation at each $n$ for 20 sets of hypothetical trials. 
We set $n=20$ subjects per condition based on this sampling to keep the predicted standard deviation below 2\% for all conditions. 

Participants were recruited through Amazon Mechanical Turk/Cloud Research, restricted to workers in the US with at least a 90\% approval rating. The experiment took $<5$ min. and participants were compensated with \$0.60. We collected data in batches until there were least 20 participants in each of 20 conditions who passed the exclusion criteria. Due to random assignment and participants sometimes dropping out, sample sizes ranged from 20-27 per condition. We analyzed data from 1723 participants (1,038 women, 665 men, 11 nonbinary, 1 agender, 1 demigirl, 1 queer, 1 genderqueer, 2 genderfluid, 1 transmasculine, 1 none, and 1 preferred not to say).

Participants were excluded if they did not complete the task and submit to mTurk, or if they did not pass the color vision test. The color vision test had two parts. Participants typed the number that they saw in each of 11 Ishihara plates (or typed ``none'' if they did not see a number). Then, participants were asked: (a) Do you have difficulty seeing colors or noticing differences between colors, compared to the average person? (b) Do you consider yourself to be color blind? Participants were excluded if they made errors on $>2$ plates or answered yes to at least one color vision question.

All participants gave informed consent and the UW--Madison IRB approved the protocol. Data, code, and color coordinates can be found at \url{https://github.com/SchlossVRL/spatial_maps}.

\begin{figure}[tb]
    \centering
    \includegraphics[scale=0.85]{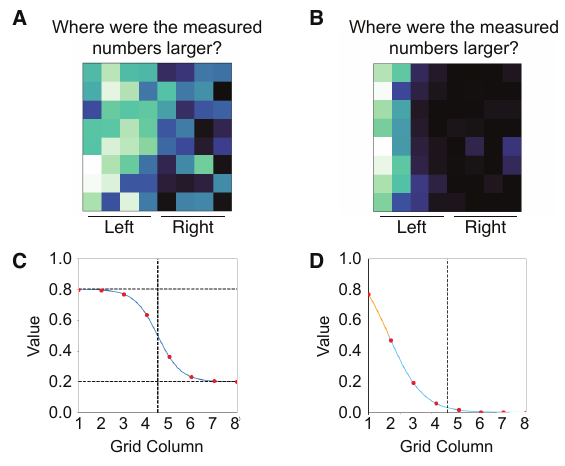} 
    \vspace{-5mm}
    \caption{Example experimental trials for the (A) unshifted and (B) shifted grid colormaps, and corresponding Sigmoid curves for the (C) unshifted and (D) shifted conditions used to sample the values at each point for generating the colormap images.}
    \label{fig:trial_curves}
    \vspace{-5mm}
\end{figure}

\textbf{Design, Displays, and Procedure.}
Participants were told that they would see colormap data visualizations representing ``measured data.'' As in \cite{mcgranaghan1989}, they were provided no details about the source of the data to avoid effects of direct associations \cite{schoenlein2023}. The maps would be displayed one at a time, and their task was to indicate whether the measured numbers were larger on the left or right side of the colormap (Fig. ~\ref{fig:trial_curves}A-B). For each colormap, one side was biased to be lighter and one was biased to be darker (left/right balanced over trials). Below each side was a horizontal line and label indicating whether the side was "left" or "right." Colormaps were presented without legends so we could probe inferred mappings directly, as in \cite{cuff1973, mcgranaghan1989, schoenlein2023}. This method is ecologically valid, as colormaps are often presented without legends in the wild \cite{christen2013}.

Participants were randomly assigned to one of 80 between-subject conditions, which included 2 granularity levels (coarse vs. fine) $\times$ 2 data distributions (dark-shifted, unshifted) $\times$ 2 background colors (white, black) $\times$ 10 color scales (Autumn, ColorBrewer Blue, ColorBrewer Red, Gray, Hot, Magma+, Mako+, Plasma, Viridis, Winter) (see Fig. \ref{fig:colormaps}). In each condition, participants judged 10 colormap visualizations in random order. The 10 colormaps were constructed from five underlying datasets, which were left/right reflected, balancing which side was darker. We also collected data for light-shifted colormaps, but focus on the unshifted and dark-shifted conditions to test our key hypothesis. We will analyze the full dataset in a subsequent paper.

\textit{\textbf{Data distribution}}. To generate the colormaps in the unshifted condition, we used a Sigmoid function \begin{math} S(x) = (L/1+e^{(K*x)})+d \end{math} where \begin{math}  \lim_{x\to 0} S(x) = 0.8 \end{math} and \begin{math}  \lim_{x\to 1} S(x) = 0.2 \end{math}  (Fig.~\ref{fig:trial_curves}C). We shifted the original Sigmoid curve along the x-axis such that \begin{math} \lim_{x\to 0.7} S(x) = 0 \end{math} and \begin{math} \lim_{x\to 0} S(x) \approx 0.8 \end{math} for the shifted conditions shown as Fig.~\ref{fig:trial_curves}D. 

\textit{\textbf{Granularity.}} The colormap visualizations with coarse granularity appeared as $8\times8$ grids (based on stimuli from \cite{schloss2019, schoenlein2023}). They were generated by sampling from a Sigmoid curve with normally-distributed noise applied. To generate the value for each cell in the grid, we first applied linear interpolation to sample 8 points \begin{math} x_i \end{math} in \begin{math} x:[0, 1] \end{math}, and calculated the corresponding \begin{math} y_i \end{math} with a Sigmoid function. After retrieving \begin{math} y_i \end{math}, which we used as means in the probability density function of the normal distribution with \begin{math} \sigma = 0.1 \end{math} to randomly sample 8 points as the column values. We clipped values that fell out of the range of [0,1] so they fit within the range. The colormaps with fine granularity were constructed as $200\times200$ grids that appeared as continuous data. They were constructed by generating a $50\times50$ grid using the same strategy as the coarse granularity above, then interpolating it to target size ($200\times200$) by applying random normal distribution with the mean of 4 adjacent point values and \begin{math} \sigma = 0.03 \end{math}. Finally, we applied median blur with a kernel (n=5) to smooth the colormaps. All colormaps were $627\times627$ px.

\textit{\textbf{Background}}. The backgrounds of the colormaps were black ([R = 0, G = 0, B = 0]) or white ([R = 255, G = 255, B = 255]), and filled the monitor display. When the background was black, the text, map border, and lines below the visualization were white. When the background was white, these components were black. 

\textit{\textbf{Color scales}}. We applied 10 different color scales to each underlying dataset. We chose color scales that fell into five different ``families'': monochromatic (``ColorBrewer Blue,'' ``ColorBrewer Red''), achromatic (``Gray,''), hue spiral with black and white end points (``Hot,'' ``Magma+,'' ``Mako+''), hue spiral with chromatic endpoints (``Plasma,'' ``Viridis''), and hue segment (``Autumn,'' ``Winter''). 

Plasma, Magma, and Viridis were created by van der Walt and Smith for the Matplotlib library in Python \cite{hunter2007}. ColorBrewer Blue and Red were created by Harrower and Brewer (2003) \cite{harrower2003}. Gray, Hot, Autumn, and Winter are native to MATLAB. ``Magma+'' and ``Mako+'' are adapted from  Magma \cite{hunter2007} and Mako (from the Seaborn library for Python \cite{waskom2021}) respectively, but we extended the light endpoints so that the lightest values were white. To make Magma+, we (1) removed the 30 darkest colors, (2) appended 20 steps to the lightest side of the scale by interpolating the lightest color (L*=97.850; a*=-9.918; b*=29.506) with white (L*=100; a*=0, b*=0), and (3) appended 10 steps to the darkest side by interpolating the darkest color (L*=8.397, a*=19.905, b*=-28.862) black (L*=0; a*=0; b*=0). To make Mako+, we (1) appended 10 steps to the lightest side by interpolating the lightest color (L*=93.383, a*=-10.979, b*=5.533) with white, and (2) 5 steps to the darkest side by interpolating the darkest color (L*=10.359, a*=11.442, b*=-10.144) with black.

\section{Results and Discussion}
\begin{figure}[tb]
    \centering
    \vspace{-5mm}
    \includegraphics[width=\columnwidth]{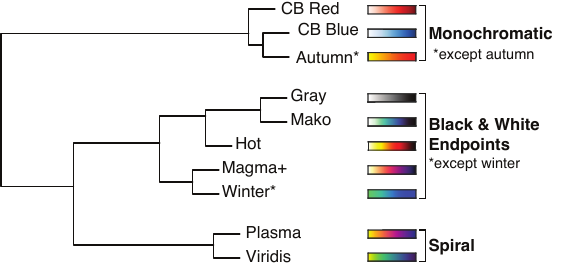} 
    \caption{Dendrogram showing hierarchical clustering of color scales according to responses for the 8 conditions within each color scale.}
    \label{fig:dendrogram}
    \vspace{-5mm}
\end{figure}

When designing this experiment, we chose color scales that fell into five different ``families'' (see Section \ref{sec:Methods}). To test whether this structure was reflected in the data, we used hierarchical clustering to group the color scales according to patterns of responses \cite{statsR}. The clustering was computed over the proportion of times participants indicated the darker side of the colormap represented larger quantities (averaged over repetitions and participants) for each of the eight conditions within each color scale: 2 shift conditions (shifted/unshifted) $\times$ 2 granularities (grid/continuous) $\times$ 2 backgrounds (white/black) (see Fig.~\ref{fig:colormaps}). Results of the hierarchical clustering are shown in Fig.~\ref{fig:dendrogram}.

The color scales fell into three main clusters, largely aligning with our initial family classification. One cluster included the monochromatic color scales (ColorBrewer Blue and ColorBrewer Red), plus Autumn. We  call this cluster \textit{Monochromatic} (\textit{Mono}) because 2/3 color scales are monochromatic, acknowledging Autumn does not fit the description. A second cluster included Gray, Mako+, Hot, Magma+, and Winter. We call this \textit{Black and white endpoints} (\textit{B\&W}) because 4/5 scales have black/white endpoints, acknowledging Winter does not fit this description. The third cluster included the hue spiral color scales with chromatic endpoints (Plasma, Viridis) so we call it \textit{Spiral}. Work is needed to understand why Autumn and Winter joined other clusters rather than forming their own as expected.

\begin{figure}[tb]
    \centering
    \includegraphics[width=\columnwidth]{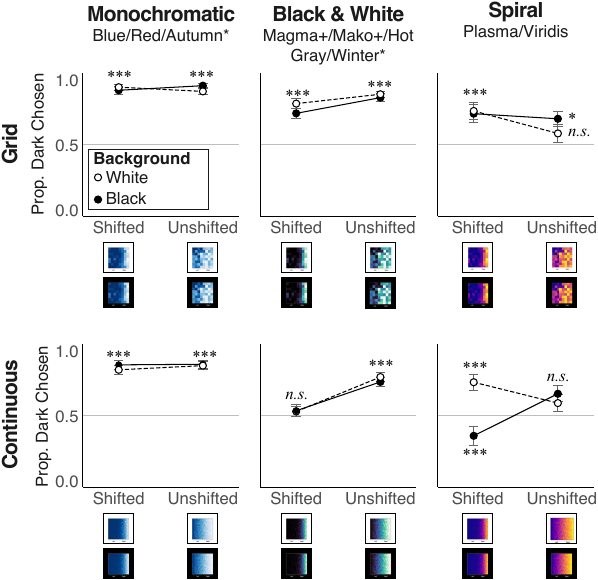} 
    \vspace{-5mm}
    \caption{Mean proportion of dark-more responses, averaged over participants and color scales within each cluster (Mono, B\&W, and Spiral). Shift is on the x-axis and background color is using mark fill (black/white marks indicate black/white backgrounds). Error bars represent +/- standard errors of the means. Significant difference from chance (Bonferroni-Holm corrected) is indicated near each point ($*** <.001$, $*<.05$, \textit{ns} = not significant. Single indicators of significance placed at one x-axis location apply for both backgrounds.} 
    \label{fig:results}
    \vspace{-5mm}
\end{figure}

Figure~\ref{fig:results} shows the mean proportion of times participants indicated the darker side of the colormap meant larger quantities, averaged over all color scales within each cluster: Mono, B\&W, Spiral (see Supplementary Material Figure~\ref{fig:line_plots_all} for results separated by color scale). The data are plotted as a function of data distribution (shifted vs. unshifted) for colormaps presented on a white background or black background, separated by granularity (grid vs. continuous). We analyzed the data using a mixed-effect logistic regression model predicting whether participants chose the lighter (0) or  darker (+1) region on each trial. The fixed effects were data distribution, background, granularity, two orthogonal contrasts coding for color scale cluster, and all possible interactions. One contrast (Scale1) compared Spiral (+2) vs. Mono (-1) and B\&W (-1) and the other (Scale2) compared B\&W (+1) with Mono (-1) with Spiral coded as 0. The model also included random by-subject intercepts.

Full model results are shown in Supplementary Material Table ~\ref{tab:model1}, with main findings summarized here. Participants were overall more likely than chance to choose the darker side  (positive intercept; $\beta=7.415$, $SE=0.38$, $z=19.722$, $p<.001$). The probability of dark-more judgments decreased for shifted colormaps ($\beta=0.818$, $SE=0.38$, $z=2.18$, $p=.029$). This effect depended on granularity, with a larger decrease in dark-more responses for shifted datasets when colormaps were continuous vs. grids (shift $\times$ granularity interaction; $\beta=1.00$, $SE=0.38$, $z=2.66$, $p=.008$). This two-way interaction was part of a larger interaction with background and Scale1 (Spiral vs. Mono and B\&W scales; $\beta=0.915$, $SE=0.34$, $z=2.72$, $p=.007$).

To understand this 4-way interaction, we conducted separate versions of the model within each color scale cluster (see Supplementary Table ~\ref{tab:model2} for the full output of each model). We also conducted intercept-only models for each condition to test whether responses for each condition in Fig. \ref{fig:results} differed from chance after applying the Holm-Bonferroni correction for multiple comparisons. Results are shown as asterisks in Fig. \ref{fig:results}, and the model output is in Supplementary Material Table~\ref{tab:intercept_tests}. 

\textbf{Monochromatic.} Only the intercept was significant in this model ($\beta=8.867$, $SE=0.47$, $z=18.850$, $p<.001$)---participants consistently made dark-more responses for all eight conditions, regardless of shift, background, or granularity (Figure \ref{fig:results}).

\textbf{Black and white endpoint.} In this cluster, the hole hypothesis predicted an interaction between data distribution and background, with reduced dark-more responses for the shifted condition on the black background due to a large region of map appearing as a hole. Overall, dark-more responses were reduced on for the shifted than unshifted condition (main effect of shift; $\beta=0.654$, $SE=0.266$, $z=2.461$, $p=.014$), but this interaction with background was not significant ($\beta=-0.015$, $SE=0.254$, $z=-.062$, $p=.951$; see Figure \ref{fig:results}).  Tests against chance indicated that participants consistently made dark-more responses for all conditions except the shifted continuous condition, where responses did not differ from chance. It is unclear why this effect of shift occurred on both white and black backgrounds, and future work will address this question.

\textbf{Spiral.} We expected responses for the Spiral color scales to be similar to the Monochromatic color scales, but they were drastically different. There was an unexpected 3-way interaction between data distribution, background, and granularity ($\beta=-2.267$, $SE=0.420$, $z=-5.402$, $p<.001$). For grid colormaps, participants made dark-more responses more often than chance for all conditions except dark background--unshifted. For continuous maps, participants made dark-more responses more often than chance for the shifted condition on the light background, but unexpectedly made \textit{light-more responses} more often than chance on the dark background (Figure \ref{fig:results}, Table \ref{tab:intercept_tests}). Thus, we found a condition in which participants consistently made light-more responses, but it was not the condition we expected. Future work is needed to understand this effect.

\section{Conclusion} 
We hypothesized that the percept of a hole in a colormap would activate the opaque-is-more bias even if the color scale did not appear to vary in opacity in isolation (Hot, Mako+, Magma+ color scales, see Black and White endpoint condition). Activation of the opaque-is-more bias would lead to reduced dark-more responses for shifted maps on a black background, relative to the same maps on a white background (hole hypothesis). In the extreme, this effect could have led to \textit{light-more} responses, inferring that lighter colors mapped to larger quantities.

Yet contrary to the hole hypothesis, background did not significantly modulate responses for the B\&W scales that had strong perceptual evidence for a hole, but did modulate responses for the Spiral scales that had weak perceptual evidence for a hole. For Spiral scales on a black background, participants were more likely than chance to make \textit{light-more} responses. Based on the Opacity Variation Index \cite{schloss2019}, Spiral color scales should not appear to vary in opacity and should not appear to have a ``hole.'' The reason for these results is unknown.  

Our findings raise new questions about the contributions of spatial configuration and granularity to inferences about colormap data visualizations. These variables have an effect on inferred mappings and they interact with background color and color scales in unexpected ways. Although future work is needed to explain these effects, our study has shown that it is crucial to account for the spatial distribution of data when considering color semantics in data visualizations. 

\acknowledgments{
We thank Kushin Mukherjee, Melissa Schoenlein, and Seth Gorelik for feedback. This project was supported by the UW--Madison Office of the Vice Chancellor for Research and Graduate Education, Wisconsin Alumni Research Foundation, McPherson Eye Research Institute, and NSF BCS-1945303 to KBS and IIS-2320920 to DAS. }


\clearpage
\balance

\bibliographystyle{abbrv-doi}

\bibliography{References.bib}


\clearpage
\appendix
\twocolumn\renewcommand*{\thesection}{S}
\counterwithin{figure}{section}
\counterwithin{table}{section}

\twocolumn\section{Supplementary Material}\label{sec:supplementary}

\begin{table}[ht!] 
\centering
  \caption{Logistic mixed-effects model predicting probability of choosing the darker side from data distribution (shift), background (bg), granularity (granular), and two color scale contrasts (scale1 and scale2, see text for details). Note: $\beta$ represents the regression coefficients, SE represents standard error, \textit{z} represents z-scores, and \textit{p} represents p-values.}
  \label{tab:model1}
  \begin{tabular}{lrrrr}
  \toprule
  \bf{Fixed Effects} & $\bf{\beta}$ & \bf{SE} & $\bf{z}$ & $\bf{p}$ 
  \\
  \midrule
  Intercept                 & $7.415$ & $0.38$ & $19.72$ & $<.001$
  \\ 
  Shift                     & $0.818$ & $0.38$ & $2.18$ & $.029$
  \\
  Scale1                    & $-1.387$ & $0.34$ & $-4.12$ & $<.001$
  \\
  Scale2                    & $-1.026$ & $0.42$ & $-2.44$ & $.015$
  \\ 
  Bg                        & $ 0.145$ & $0.38$ & $0.39$ & $.699$
  \\
  Granular                  & $-1.271$ & $0.38$ & $-3.38$ & $<.001$
  \\ 
  Shift*Scale1              & $-0.128$ & $0.34$ & $-0.38$ & $.704$
  \\
  Shift*Scale2              & $0.693$ & $0.42$ & $1.65$ & $.099$
  \\
  Shift*Bg                  & $-0.485$ & $0.38$ & $-1.29$ & $.197$
  \\
  Scale1* Bg                & $0.524$ & $0.34$ & $1.56$ & $.120$
  \\
  Scale2*Bg                 & $0.011$ & $0.42$ & $0.03$ & $.978$
  \\
  Shift*Granular            & $1.000$ & $0.38$ & $2.66$ & $.008$
  \\
  Scale1*Granular           & $-0.411$ & $0.34$ & $-1.22$ & $.222$
  \\
  Scale2*Granular           & $-0.590$ & $0.42$ & $-1.40$ & $.160$
  \\
  Bg*Granular               & $0.283$ & $0.38$ & $0.75$ & $.452$
  \\
  Shift*Scale1*Bg           & $-1.098$ & $0.34$ & $-3.26$ & $.001$
  \\
  Shift*Scale2*Bg           & $0.139$ & $0.42$ & $0.33$ & $.741$
  \\
  Shift*Scale1*Granular     & $0.500$ & $0.34$ & $1.49$ & $.137$
  \\
  Shift*Scale2*Granular     & $0.516$ & $0.42$ & $1.23$ & $.220$
  \\
  Shift*Bg*Granular         & $-0.144$ & $0.38$ & $-0.38$ & $.702$
  \\
  Scale1*Bg*Granular        & $0.915$ & $0.34$ & $2.72$ & $.007$
  \\
  Scale2*Bg*Granular        & $-0.098$ & $0.42$ & $-0.23$ & $.815$
  \\
  Shift*Scale1*Bg*Granular  & $-0.811$ & $0.34$ & $-2.41$ & $.016$
  \\
  Shift*Scale2*Bg*Granular  & $0.116$ & $0.42$ & $0.27$ & $.784$
  \\
  \bottomrule
  \end{tabular}
\end{table}

\begin{table}[H]
\centering
  \caption{Mixed-effects logistic regression model predicting probability of choosing the darker side from data distribution (shift), granularity (granular), background (bg), and their interaction for each cluster of color scales (Mono, B\&W, Spiral).}
  \label{tab:model2}
  \begin{tabular}{lrrrr}
  \toprule
  \textbf{Monochromatic}
  \\
  \midrule
\bf{Condition} &$\bf{\beta}$ & \bf{SE} & $\bf{z}$ & $\bf{p}$

\\
 Intercept & $8.867$ & $0.470$ & $18.850$ & $<.001$
 \\
 Shift          & $-0.071$ & $0.277$ & $-0.257$ & $.797$
 \\
 Bg.            & $-0.134$ & $0.277$ & $-0.485$ & $.628$
 \\
 Granular       & $-0.274$ & $0.277$ & $-0.989$ & $.323$
 \\
 Shift*Bg       & $-0.143$ & $0.277$ & $-0.515$ & $.606$
 \\
 Shift*Granular & $0.052$ & $0.277$ & $0.187$ & $.852$
 \\
 Bg*Granular. & $-0.032$ & $0.277$ & $-0.118$ & $.906$
 \\
 Shift*Bg.*Granular & $0.099$ & $0.277$ & $0.357$ & $.721$
 \\
  \bottomrule
  \\
\textbf{B\&W Endpoints}
  \\
  \midrule
\bf{Condition} & $\bf{\beta}$ & \bf{SE} & $\bf{z}$ & $\bf{p}$
\\
 Intercept & $8.908$ & $0.460$ & $19.347$ & $<.001$
 \\
 Shift & $0.654$ & $0.266$ & $2.461$ & $.014$
 \\
 Bg. & $0.087$ & $0.254$ & $0.343$ & $.732$
 \\
 Granular & $-0.730$ & $0.268$ & $-2.729$ & $.006$
 \\
 Shift*Bg & $-0.015$ & $0.254$ & $-0.062$ & $.951$
 \\
 Shift*Granular & $0.382$ & $0.265$ & $1.440$ & $.149$
 \\
 Bg*Granular & $-0.049$ & $0.254$ & $-0.192$ & $.847$
 \\
 Shift*Bg*Granular & $0.140$ & $0.254$ & $.551$ & $0.582$
 \\
  \bottomrule
\\
\textbf{Spiral}     
\\ 
\midrule    
\bf{Condition} & $\bf{\beta}$ & \bf{SE} & $\bf{z}$ & $\bf{p}$
\\
 Intercept          & $6.490$ & $0.585$ & $11.097$ & $<.001$
 \\
 Shift              & $1.543$ & $0.460$ & $3.352$ & $<.001$
 \\
 Bg                 & $2.089$ & $0.435$ & $4.802$ & $<.001$
 \\
 Granular           & $-2.335$ & $0.416$ & $-5.609$ & $<.001$
 \\
 Shift*Bg           & $-2.529$ & $0.408$ & $-6.201$ & $<.001$
 \\
 Shift*Granular  & $2.206$ & $0.421$ & $5.232$ & $<.001$
 \\
 Bg*Granular        & $2.375$ & $0.417$ & $5.693$ & $<.001$
 \\
 Shift*Bg*Granular & $-2.267$ & $0.420$ & $-5.402$ & $<.001$
 \\
 \bottomrule
  
  \end{tabular}
\end{table}

\begin{figure*}[ht!] 
    \centering
    \includegraphics[scale=1]{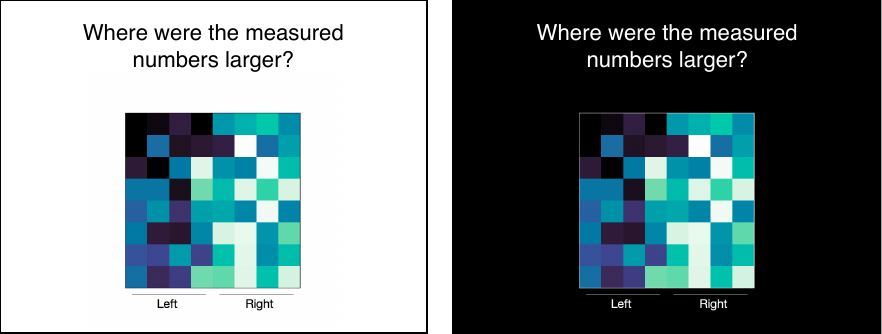} 
    \caption{Examples of trials on the white background (left) or black background (right) conditions. The colormaps shown in these trials are coarse granularity unshifted maps generated using the Mako+ color scale.} 
    \label{fig:example_trials}
    \vspace{-5mm}
\end{figure*}

\begin{figure*}[tb]
    \centering
    \includegraphics[scale=.88]{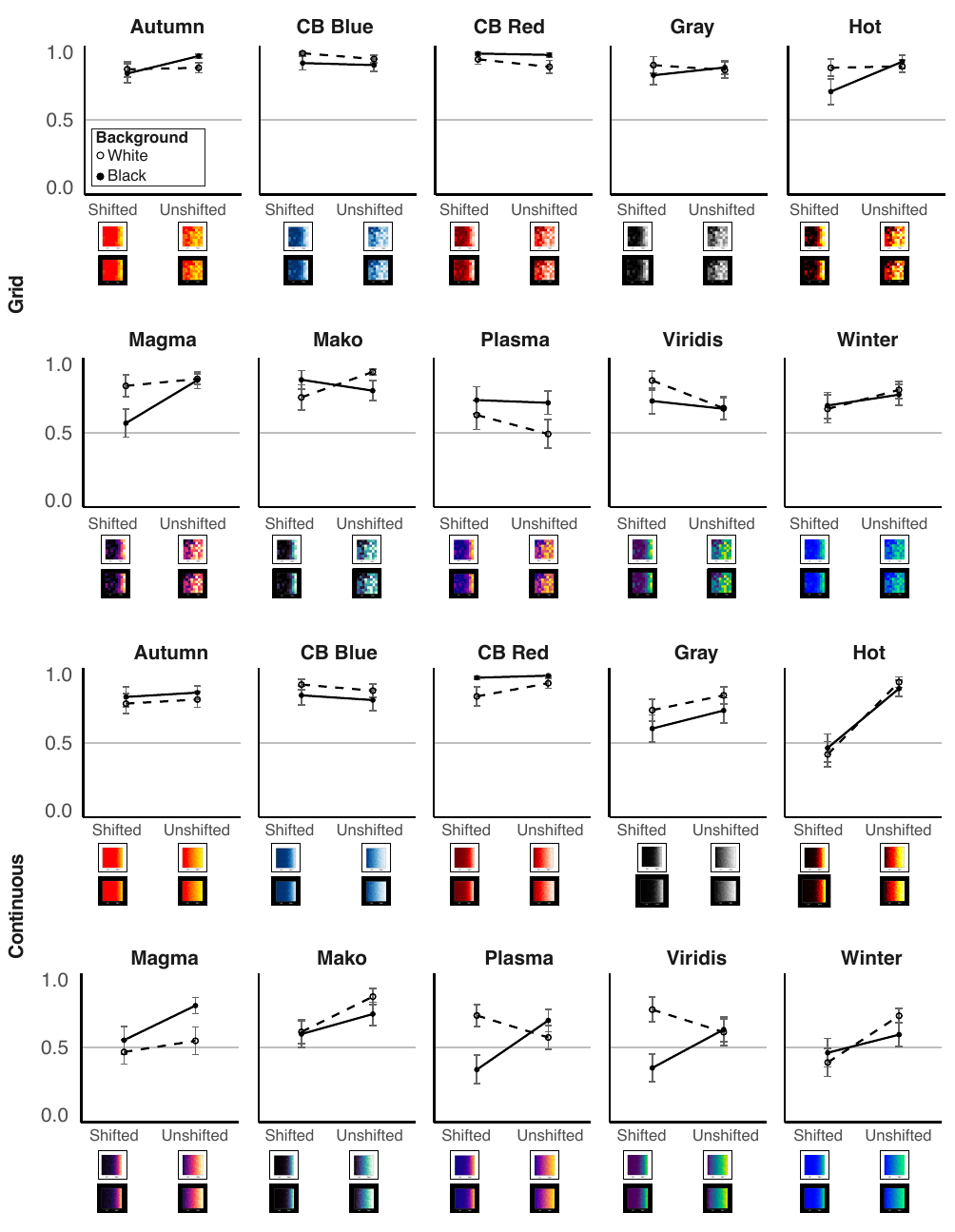} 
    \caption{Mean proportion of dark-more responses, averaged over participants and separated by color scale. Shift is coded on the x-axis and background color is coded using mark fill (black/white marks indicate black/white background, respectively). Error bars represent +/- standard errors of the means. } 
    \label{fig:line_plots_all}
    \vspace{-5mm}
\end{figure*}

\begin{table*}[t]
\centering
  \caption{Mixed-effects logistic regression models predicting probability of choosing the darker side from intercept only for each of 24 conditions.}
  \label{tab:intercept_tests}
  \begin{tabular}{lrrrr}
  \toprule
\bf{Condition} & $\bf{\beta}$ & \bf{SE} & $\bf{z}$ & $\bf{p}$
 \\
 Shift/Grid/Spiral-BW/White bg. & $11.99$ & $1.810$ & $6.624$ & $<.001$ 
 \\
 Shift/Grid/Spiral-BW/Black bg. & $12.220$ & $1.822$ & $6.706$ & $<.001$ 
 \\
 Unshift/Grid/Spiral-BW/Black bg. & $8.926$ & $0.984$ & $9.068$ & $<.001$ 
 \\ 
 Shift/Continuous/Mono/White bg. & $10.33$ & $1.280$ & $8.066$ & $<.001$
 \\
 Unshift/Continuous/Spiral-BW/White bg. & $8.092$ & $1.011$ & $8.003$ & $<.001$ 
 \\
 Unshift/Continuous/Spiral-BW/Black bg. & $7.983$ & $1.032$ & $7.736$ & $<.001$
 \\
 Shift/Continuous/Mono/Black bg. & $9.825$ & $1.422$ & $6.908$ & $<.001$
 \\
 Shift/Grid/Spiral/Black bg. & $12.220$ & $1.822$ & $6.706$ & $<.001$
 \\
 Shift/Continuous/Spiral/White bg. & $11.99$ & $1.810$ & $6.624$ & $<.001$
 \\
 Shift/Grid/Mono/White bg. & $9.405$ & $1.426$ & $6.597$ & $<.001$
 \\
 Unshift/Continuous/Mono/Black bg. & $8.983$ & $1.362$ & $6.594$ & $<.001$
 \\
 Shift/Grid/Mono/Black bg. & $9.860$ & $1.498$ & $6.582$ & $<.001$
 \\
 Shift/Continuous/Spiral/Black bg. & $-10.694$ & $1.662$ & $-6.434$ & $<.001$ 
 \\
 Shift/Grid/Spiral/White bg. & $11.069$ & $1.759$ & $6.291$ & $<.001$ 
 \\
 Unshift/Grid/Spiral-BW/White bg. & $6.410$ & $1.187$ & $5.400$& $<.001$
 \\
 Unshift/Grid/Mono/Black bg. & $7.966$ & $1.477$ & $5.393$ & $<.001$
 \\
 Unshift/Grid/Mono/White bg. & $6.406$ & $1.500$ & $4.269$ & $<.001$
 \\
 Unshift/Continuous/Mono/White bg. & $5.651$ & $1.332$ & $4.242$ & $<.001$
 \\
 Unshift/Grid/Spiral/Black bg. & $2.283$ & $ 0.732$ & $3.117$ & $0.011$
 \\
 Unshift/Continuous/Spiral/Black bg. & $1.978$ & $0.821$ & $2.408$ & $0.080$
 \\
 Unshift/Continuous/Spiral/White bg. & $1.188$ & $0.848$ & $1.401$ & $0.644$
 \\
 Shift/Continuous/Spiral-BW/Black bg. & $1.305$ & $0.960$ & $1.359$ & $0.644$ 
 \\
 Unshift/Grid/Spiral/White bg. & $1.284$ & $1.297$ & $0.990$ & $0.644$
 \\
 Shift/Continuous/Spiral-BW/White bg. & $1.337$ & $1.465$ & $0.913$ & $0.644$
 \\
  \bottomrule
  \end{tabular}
\end{table*}

\end{document}